\documentclass[11pt]{article}
\usepackage{amssymb,amsmath}

\begin{document} 
\title{Equivalence between two-dimensional cell-sorting 
and one-dimensional generalized random walk, \\
-- spin representations of generating operators--. } 
\author{Kazuhiko MINAMI}
\date{6/30/2011}
\maketitle
\abstract{ 
The two-dimensional cell-sorting problem is found to be mathematically equivalent 
to the one-dimensional random walk problem with pair creations and annihilations, 
i.e. the adhesion probabilities in the cell-sorting model 
relate analytically to the expectation values in the random walk problem. 
This is an example demonstrating that two completely different biological systems 
are governed by a common mathematical structure. 
This result is obtained 
through the equivalences of these systems with lattice spin models. 
It is also shown that 
arbitrary generation operators can be written by the spin operators, 
and hence all biological stochastic problems can in principle be analyzed 
utilizing the techniques and knowledge previously obtained 
in the study of lattice spin systems. }
\\
\vspace{2.4cm}

\vspace{0.3cm}

\noindent
Graduate School of Mathematics, Nagoya University,
Nagoya, 464-8602, JAPAN

\noindent
minami@math.nagoya-u.ac.jp

\newpage

\section{Introduction}
Lattice models are important in biology 
when one needs to introduce a spatial structure. 
It has been shown that 
lattice models are often equivalent to spin models, 
which are lattice models for magnetic materials. 
Spin models have long been studied as a purely statistical mechanical subject, 
and various models on various lattices with different types of interactions 
have been investigated in detail. 
Approximation methods and techniques for numerical calculations 
have been extensively developed.  
Some models have been solved and the analytic expressions for physical quantities 
such as energy, specific heat, susceptibility, or correlation functions have been obtained. 
Some of the solvable models are known to be equivalent to each other, 
i.e. analytic relations between the quantities of the models 
are found under some specific relations of parameters. 

In this study, it is shown that 
models for biological systems (or ecosystems, or organismic systems) 
are sometimes governed by a common mathematical structure,  
even though the models seem to differ completely from other.  
We show that the two-dimensional cell-sorting 
is equivalent to a generalized one-dimensional random walk 
in the following manner. 
The two-dimensional cell-sorting model on a square grid 
is equivalent to the two-dimensional Ising model 
which is one of the most typical statistical mechanical models for magnetism. 
Prior study have shown 
the two-dimensional Ising model is equivalent to the one-dimensional XY model, 
the Hamiltonian of which can be interpreted as the generation matrix 
of the random walk problem with pair creations and annihilations. 

In Section 2, 
the spin models are briefly reviewed. 
In Section 3, 
a cell-sorting model is reviewed and its equivalence to the Ising model is considered.  
In Section 4, 
the analytic relations between the adhesion probabilities in the cell-sorting problem 
on a two-dimensional square grid 
and the expectation values in the random walk problem 
with pair creations and pair annihilations are derived. 

In Section 5, 
relations between the stochastic movements, particularly those of the molecular motors,  
and spin Hamiltonians are considered. 
Molecular motors are protein molecules that are vital to biological motion 
especially for  internal material transport such as muscle contraction, bacterial motion, 
cell division, intracellular transport along the axons of nerve cells, and genomic transcription.
Several important families of motor proteins 
such as ribosomes, kinesins, dyneins, myosins and RNA polymerase 
have been identified. 
They move along periodically structured molecular tracks  
and show stochastic movements on an one-dimensional template.  
In this sutudy, 
I will consider the random walk problem with hard-core interaction, 
the movements of ribosomes on a mRNA, 
and the movements of kinesins.  
Finally I will show that 
generation rules for stochastic movements appearing in biological problems 
can always be represented by the Hamiltonians of spin models.

\section{Lattice spin models and their equivalences}
{\bf Lattice spin models:} 
Magnetic materials are composed of atoms 
that have their own magnetic moment. 
The atoms interact with each other energetically,  
and these interactions are completely quantum mechanical in nature. 
Spin models are the mathematical models for magnets 
in which spin operators are assigned to every lattice point and interact with each other.  
Quantum spin is a kind of angular momentum 
and defined through the following commutation relations: 
\begin{eqnarray}
[s_l^x, s_l^y]=is_l^z, \hspace{0.6cm}
[s_l^y, s_l^z]=is_l^x, \hspace{0.6cm}
[s_l^z, s_l^x]=is_l^y,
\label{comm}
\end{eqnarray}
where $[A, B]=AB-BA$, $i^2=-1$, and $s_l^x$, $s_l^y$, $s_l^z$ 
are the $x$, $y$, $z$ components of the spin operator at site $l$, respectively. 

Assume the spin at site $i$ and the spin at site $j$ energetically interact with each other. 
Let the interaction energy be   
$-J(s_i^x s_j^x+s_i^y s_j^y+\Delta s_i^z s_j^z)$, 
where $J$ is the coupling constant and $\Delta$ is the anisotropic parameter. 
The Hamiltonian of the system is written as 
\begin{eqnarray}
H=-J\sum_{\langle i,j\rangle}[s_i^x s_j^x+s_i^y s_j^y+\Delta s_i^z s_j^z],
\label{XXZ}
\end{eqnarray}
which is the sum of the interaction energies 
over all the spin pairs $\langle i,j\rangle$ that interact each other. 
The eigenvalues of $H$ are the energies which can be realized in this system. 
The model (\ref{XXZ}) is called the XXZ model, 
its isotropic case $\Delta=1$ is called the Heisenberg model.

Spin operators are classified by the spin magnitude $S$. 
Possible values of $S$ are $0$, $1/2$, $1$, $3/2$ 
and $m/2$ where $m$ is a natural number. 
Let us consider the most typical case: $S=1/2$.  
In this case, there are two eigenstates of $s_i^z$ with the eigenvalues $1/2$ and $-1/2$.  
We call them the spin up state $|+\rangle_i$ and the spin down state $|-\rangle_i$, respectively.  
Generally there exist $n=2S+1$ eigenstates of the spin operator $s_i^z$
when its magnitude is equal to $S$. 

Then we find that 
the energy contribution from the last term in (\ref{XXZ}) 
is obtained from the product of the eigenvalues of interacting two spins. 
We call this factor the Ising interaction. 
Spin model with only the Ising interaction is called the Ising model. 

The one-dimensional Ising model 
\begin{eqnarray}
H=-J\sum_{i=1}^Ns_i^z s_{i+1}^z
\label{Ising}
\end{eqnarray}
was first solved by Ising (1925). 
When we consider the Ising model with $N$ sites, 
there are $2^N$ possible configurations of spin up and down states, 
because two states are possible for each site $i$ $(i=1, 2, \ldots, N)$. 
The total energy is a function of the configurations. 
Thus the energy can be written as $E_k$ $(k=1, 2, \ldots, 2^N)$ 
where each $k$ denotes a configuration. 
Following the statistical mechanics, 
the probability to find a configuration with an energy $E_k$ is $\exp(-E_k/k_{\rm B}T)/Z$, 
where $T$ is the temperature, $k_{\rm B}$ is the Boltzmann constant, 
and $Z=\sum_{k=1}^{2^N}\exp(-E_k/k_{\rm B}T)$ is the normalization factor 
called the partition function. 
Then the expectation value of the energy, for example, is obtained as 
\begin{eqnarray}
\langle E\rangle=\sum_{k=1}^{2^N}E_k\exp(-\beta E_k)/Z,  
\hspace{0.6cm}\beta=\frac{1}{k_{\rm B}T},
\nonumber
\end{eqnarray}
which can also be written as 
\begin{eqnarray}
\langle E\rangle
=\frac{1}{Z}\frac{\partial}{\partial(-\beta)}\sum_{k=1}^{2^N}\exp(-\beta E_k)
=\frac{\partial}{\partial(-\beta)}\log Z. 
\nonumber
\end{eqnarray}
We then introduce the Gibbs free energy $F$ by the relation $-\beta F=\log Z$.  
The expectation value of the total energy $E$, 
and other quantities like magnetization, specific heat and susceptibility, 
are obtained as the derivatives of $F$. 
Models are usually said to be solved 
when their Gibbs energies are analytically obtained in a closed form. 

The Gibbs free energy of the one-dimensional Ising model have also been obtained 
in a sophisticated way in introducing the transfer matrix method 
(Kramers and Wannier 1941; Kubo 1943). 
Let $m_i$ be the eigenvalue of $s_i^z$. 
The partition function of this model is written as 
\begin{eqnarray}
Z&=&\sum_{k=1}^{2^N}\exp[-\beta E_k]\nonumber\\ 
&=&\sum_{m_1=\pm\frac{1}{2}}\sum_{m_2=\pm\frac{1}{2}}\cdots \sum_{m_N=\pm\frac{1}{2}}
\exp[\beta J (m_1m_2+m_2m_3+\cdots m_Nm_1)] \nonumber\\ 
&=&\sum_{m_1=\pm\frac{1}{2}}\sum_{m_2=\pm\frac{1}{2}}\cdots \sum_{m_N=\pm\frac{1}{2}}
\exp(\beta J m_1m_2)\exp(\beta J m_2m_3)\cdots \exp(\beta J m_Nm_1), \nonumber
\end{eqnarray}
where the periodic boundary condition $m_{N+1}=m_1$ is assumed. 
When one introduce $2\times 2$ matrix 
$(V)_{mm'}=\exp(\beta J mm')$, 
i.e. $\displaystyle (V)_{\frac{1}{2},\frac{1}{2}}=(V)_{-\frac{1}{2},-\frac{1}{2}}=\exp(+\beta J /4)$ and 
$\displaystyle (V)_{\frac{1}{2},-\frac{1}{2}}=(V)_{-\frac{1}{2},\frac{1}{2}}=\exp(-\beta J /4)$, 
the partition function is written as 
\begin{eqnarray}
Z&=&\sum_{m_1=\pm\frac{1}{2}}\sum_{m_2=\pm\frac{1}{2}}\cdots \sum_{m_N=\pm\frac{1}{2}}
(V)_{m_1m_2}(V)_{m_2m_3}\cdots (V)_{m_Nm_1}\nonumber\\ 
&=&\sum_{m=\pm\frac{1}{2}} (V^N)_{mm}\nonumber\\  
&=&{\rm Tr}\: V^N
=\lambda_1^N+\lambda_2^N
=\lambda_1^N[1+(\frac{\lambda_2}{\lambda_1})^N], \nonumber
\end{eqnarray}
where $\lambda_1$ and $\lambda_2$ are the eigenvalues of the matrix $V$: 
$\lambda_1=2\cosh(\beta J /4)$, $\lambda_2=2\sinh(\beta J /4)$. 
The free energy per site in the thermodynamic limit is 
\begin{eqnarray}
f
=\lim_{N\to\infty}\frac{1}{N} F  
=\lim_{N\to\infty}\frac{1}{N} \frac{1}{-\beta}\log[\lambda_1^N(1+(\frac{\lambda_2}{\lambda_1})^N)]
=\frac{1}{-\beta}\log\lambda_1. 
\nonumber
\end{eqnarray}
Effects coming from the boundary vanish 
when one take the thermodynamic limit $N\to\infty$. 
Hence the free energy is obtained 
if the maximum eigenvalue of the transfer matrix $V$ is obtained. 

When we consider the two-dimensional Ising model 
on a square lattice with the size $M\times N$, 
we have to introduce a $2^M\times 2^M$ transfer matrix. 
This huge matrix  have been diagonalized and the free energy have been obtained 
(Onsager 1944). 

All the spin operators $s_i^z$ in the Ising Hamiltonian commute with each other: 
$[s_i^z, s_j^z]=0$. 
Then the system is called a classical model because there exist no quantum effect 
coming from non-commutativity of operators.

Next let us introduce $s_l^{\rm \pm}=s_l^x\pm is_l^y$. 
From the commutation relation (\ref{comm}), 
it is derived that $s_l^{\rm +}$ maps the spin down state to the spin up state, 
$s_l^{\rm -}$ maps the spin up state to the spin down state, 
and otherwise it works as the zero operator: 
\begin{eqnarray}
s_l^{\rm +}|-\rangle_l=|+\rangle_l, \hspace{0.6cm} 
s_l^{\rm -}|+\rangle_l=|-\rangle_l, \hspace{0.6cm}
s_l^{\rm +}|+\rangle_l=0,\hspace{0.6cm} s_l^{\rm -}|-\rangle_l=0. 
\nonumber
\end{eqnarray}
The sum of the first two terms in (\ref{XXZ}) 
is equal to $(s_i^{\rm +}s_j^{\rm -}+s_i^{\rm -}s_j^{\rm +})/2$, 
and this term transfers the spin up state from site $j$ to $i$, 
or from site $i$ to $j$: 
\begin{eqnarray}
s_i^{\rm +}s_j^{\rm -}|-\rangle_i|+\rangle_j=|+\rangle_i|-\rangle_j, \hspace{0.6cm} 
s_i^{\rm -}s_j^{\rm +}|+\rangle_i|-\rangle_j=|-\rangle_i|+\rangle_j.   
\nonumber
\end{eqnarray}
Contributions coming from the other states are zero: 
$s_i^{\rm +}s_j^{\rm -}|+\rangle_i|-\rangle_j=s_i^{\rm +}s_j^{\rm -}|+\rangle_i|+\rangle_j=s_i^{\rm +}s_j^{\rm -}|-\rangle_i|-\rangle_j=0$ and 
$s_i^{\rm -}s_j^{\rm +}|-\rangle_i|+\rangle_j=s_i^{\rm -}s_j^{\rm +}|+\rangle_i|+\rangle_j=s_i^{\rm -}s_j^{\rm +}|-\rangle_i|-\rangle_j=0$. 
Thus the sum of the first two terms in  (\ref{XXZ}) is the two-body flip operation, 
in which $+$ moves from the right to the left, or from the left to the right. 
Therefore the Hamiltonian (\ref{XXZ}) is the sum of the two-body flips 
and the products of the eigenvalues of interacting two spins. 

The spin model interacting via $s_i^xs_j^x$ and $s_i^ys_j^y$,  
\begin{eqnarray}
H=-J\sum_{\langle i,j\rangle}[(1+\gamma)s_i^xs_j^x+(1-\gamma)s_i^ys_j^y]
-h\sum_{i}s_i^z, 
\label{XY}
\end{eqnarray}
is called the XY model, 
where $\gamma$ is the anisotropy parameter. 
The factor $h$ in the last term is an external magnetic field applied to the $z$ direction, 
which will become important to consider the equivalence. 
The interaction is a sum of the two-body flips 
and hence it is not trivial to find the eigenstates,  
in contrast with the fact that the eigenstates of the Ising model is obtained immediately. 
In the one-dimensional case, the XY model   
have been solved exactly (Lieb et al. 1961; Katsura 1962; Niemaijer 1967). 

The Ising model with an external field applied to the $x$ direction, $-h^x\sum_i s_i^x$,  
is called the transverse Ising model. 
The transverse susceptibility at $h^x=0$ for the one-dimensional case with $S=1/2$ 
have been exactly calculated (Fisher 1960). 
The transverse susceptibility at $h^x=0$ for the two-dimensional case with $S=1/2$
have also been obtained (Fisher 1963). 
The exact free energy for the one-dimensional case 
have been obtained (Katsura 1962; Pfeuty 1970). 
The transverse term is expressed as $s_i^x=(s_i^{\rm +}+s_i^{\rm -})/2$, 
and thus it represents independent one-body flip of each spin 
with the probability proportional to $h^x$. 
The transverse susceptibility at $h^x=0$ of the one-dimensional transverse Ising model 
with arbitrary spin magnitude $S$ have been exactly obtained (Minami 1996).  
The susceptibility and the specific heat 
for general Ising type interactions $(s_i^z)^m(s_j^z)^n$ 
have also been exactly obtained (Minami 1998). 

The Hamiltonians of the XY model and the transverse Ising model 
are the sums of operators which do not always commute with each other. 
Non-commutatibity of operators induce quantum effects, 
and these models are typical examples of the quantum spin system. \\

\noindent
{\bf Equivalences:} 
It has been derived that 
some of these spin models are equivalent to each other in the following sense (Suzuki 1971). 
The two-dimensional Ising model is solved using the transfer matrix $V$, 
where the free energy is obtained from the maximum eigenvalue of $V$. 
Let $|\phi_0 \rangle$ be the eigenstate of $V$ corresponding to the maximum eigenvalue. 
With an appropriate choice of parameters, 
the Hamiltonian $H$ of the one-dimensional XY model 
commutes with the transfer matrix $V$ of the two-dimensional Ising model: 
$[H, V]=HV-VH=0$.  
Hence these models have a common set of eigenstates: 
$H$ and $V$ can be diagonalized simultaneously. 
In particular, 
the eigenstate $|\phi_0 \rangle$ corresponding to the maximum eigenvalue of $V$ 
and the eigenstate for the smallest eigenvalue of the one-dimensional XY model 
are the same. 
This structure implies that 
the expectation values in the two-dimensional Ising model 
and the expectation values in the lowest energy state (i.e. the ground state) 
of the one-dimensional XY model analytically relate to each other. 

The two-dimensional Ising model and the one-dimensional XY model 
are equivalent to each other,  
where an anisotropic limit of the XY model with an external field 
is the transverse Ising model. 
The energy of the two-dimensional Ising model 
is determined from the simple product of the eigenvalues of interacting two spins. 
The interaction is simple but the model is not easy to solve 
because it is defined on the two-dimensional square grid. 
The interaction of the one-dimensional XY model 
is the two-body flip and it is not trivial to find the eigenstates, 
though the lattice is a one-dimensional chain and simpler than the square lattice. 
The interaction of the transverse Ising model is simple Ising interaction, 
however there exists an external magnetic field in the $x$ direction, 
which does not commute with the Ising interaction. 

It is derived that the correlation functions in the two-dimensional Ising model 
and those in the lowest energy state of the one-dimensional XY model 
satisfy the following relation: 
\begin{eqnarray}
\langle s^z_{ij}s^z_{ik}\rangle_{\rm 2D \:Ising}
=\cosh^2K_1^*\langle s^x_js^x_k\rangle_{\rm 1D \:XY}
-\sinh^2K_1^*\langle s^y_js^y_k\rangle_{\rm 1D \:XY}. 
\label{corr1}
\end{eqnarray}
The parameters are assumed to satisfy 
$\cosh 2K_1^*=1/\gamma$, $\tanh 2K_2=(1-\gamma^2)^{1/2}/h$, and 
$\sinh 2K_i\sinh 2K_i^*=1$, $K_i=\beta J_i\;(i=1, 2)$, 
where $J_1$ and $J_2$ are the vertical and horizontal coupling constants 
of the square lattice Ising model, respectively   
(Suzuki 1971, in which the Ising model is written in terms of the Pauli operator 
$\sigma^x_j$ where $s^x_j=\sigma^x_j/2$).  
The expectation value of a quantity $Q$ in the state $|\phi \rangle$ 
is calculated using the operator ${\hat Q}$ which corresponds to $Q$, 
and using the expansion $|\phi \rangle=\sum_k c_k|k \rangle$ as 
\begin{eqnarray}
\langle Q\rangle=\langle \phi |{\hat Q}|\phi \rangle
=\sum_{kk'} c_{k'}^*c_k\langle k' |{\hat Q}|k \rangle,
\label{innerproduct}
\end{eqnarray}
where $c_{k'}^*$ is the complex conjugate of $c_{k'}$ and
$\langle k' |$ is the dual state of $|k' \rangle$.  

Many other examples of equivalences of lattice spin models have been investigated. 
The six-vertex model, which is a two-dimensional lattice model, 
is equivalent to the one-dimensional XXZ model. 
The eight-vertex model is equivalent to the one-dimensional XYZ model. 
General formula for equivalences 
between $d$-dimensional classical systems and $(d+1)$-dimensional quantum systems 
have also been obtained (Suzuki 1976).

\section{A model for cell-sorting and equivalence to the Ising model}
{\bf Cell sorting problem:} 
Among the processes involved in the formation of an animal, 
one of the most important phenomena 
is the self-rearrangement of cells leading to the formation of functional structures. 
Starting from a random mixture of cells from different origins,
the cells reassemble themselves, 
begin to form clusters of cells of the same type,  
and simulate their normal histological patterns. 
These movements are directed, spontaneous, 
and proceed in the absence of external forces. 
This pattern formation phenomenon is known as cell-sorting. 

The mechanisms that determine why cells adhere to one another, 
i.e. the forces that drive cell movement during the relevant processes, 
are an important area of research.
Steinberg made two assumptions to explain certain kinds of cell rearrangements. 
He assumed that cell-sorting required spontaneous progressions 
of motile and mutually adhesive cells 
to configurations that have minimum adhesive free energy 
(Steinberg 1962a; b; c; 1963; 1970) .
His assumptions are 
i) any contact between cells has an adhesion energy depending on the cell types, and 
ii) cells are mobile and can reach a global energy minimum configuration independent of their initial condition.  
These assumptions indicate that 
differential intercellular adhesion and random movement of cells 
are the basic mechanisms of this self-organizing phenomenon. 

This differential adhesion hypothesis has been checked 
against various spatial restrictions, and various additional assumptions 
on the movement of cells have been developed.  
The cell assumed to move on either a two-dimensional square grid or a three-dimensional cube, 
has been investigated analytically, tested through numerical simulations, 
and applied to real systems 
(Goel et al. 1970; Goel and  Leith 1970; Leith and Goel 1971; Goel  and  Rogers 1978; Rogers and  Goel 1978; Mochizuki et al. 1996; 1998; Mochizuki 2002). 
Cell movement on a hexagonal grid has also been considered 
(Antonelli et al. 1973; 1975; Rogers and Sampson 1977). 
Cells are represented by not only hexagons but also general $n$-gons
(Matela and  Fletterick 1979; 1980),  
Voronoi polygons 
(Sulski et al. 1984), 
or polygonal cells
(Graner and Sawada 1993). 
Cells are represented by the large-Q Potts model, in which each cell can take Q internal states 
(Graner and Glazier 1992; Glazier and Graner 1993, Nakajima and Ishihara 2011).
A viscous liquid model with interfacial tension was considered 
(Gordon et al. 1972). 
Steiberg's theory was modified using dynamical equations of a molecular nature, 
and cell-sorting was found to occur in a near-liquid state 
(Greenspan 1981). 
A continuous mathematical model was proposed to analyze cell-sorting in Dictyostelium discoideum 
(Umeda 1989; Umeda and Inouye 1999; 2004). 

In the present study, 
I concentrate on the model  introduced by Mochizuki et al. (1996) 
on the two-dimensional square lattice. 
This model is directly equivalent to the Ising model, 
which has been exactly analyzed as a model for magnetic materials. 
\\

\noindent
{\bf Model and the equivalence:} 
Let us consider two kinds of cells distinguished by color: black and white. 
The cells  are assumed to form a regular square lattice. 
Let $\lambda_{\rm BB}$ be the strength of adhesion per cell contact 
between black and black cells, 
and $\lambda_{\rm WW}$ and $\lambda_{\rm BW}$ be the strength between white and white, and black and white cells, respectively. 
One can estimate the total adhesion $\Lambda_k$ 
when the configuration $k$ of the black and white cells is known:  
$\Lambda_k$ is the sum of all the strength of adhesion between cells. 
Cells exchange their positions between nearest neighbors. 
Let $m$ be the rate of exchange of the location. 
It is assumed the probability of finding a configuration 
having total strength equal to $\Lambda_k$ 
is proportional to $\exp(\Lambda_k/m)$. 

Let us introduce the differential adhesion 
$A=\lambda_{\rm BB}+\lambda_{\rm WW}-2\lambda_{\rm BW}$. 
The total adhesion $\Lambda_k$ changes its value by $A$ as an unit 
when two cells change their locations, 
and hence the system is controlled not through the energies 
$\lambda_{\rm BB}$, $\lambda_{\rm WW}$ and $\lambda_{\rm BW}$ independently, 
but through the differential adhesion $A$. 
Typical equilibrium configurations are calculated by Mochizuki et al.,  
where one can find a pattern 
in which black and white cells appear situated like a checker-board for $A/m=-2$,  
the pattern seems almost random when $A/m=0$, 
and it appears segregated when $A/m$ is positive and large.

This model of cell-sorting is equivalent to the Ising model 
when $\lambda_{\rm BB}=\lambda_{\rm WW}$. 
One can introduce a direct correspondence 
in which a black cell at site $i$ corresponds to the spin up state $|+\rangle_i$, 
and a white cell at site $i$ corresponds to the spin down state $|-\rangle_i$. 
The  interaction energy between two up states, or two down states, 
is equal to $-J/4$,  
and that between up and down states is equal to $+J/4$. 
Let us introduce a constant $J_0$ and set 
\begin{eqnarray}
\lambda_{\rm BB}=\lambda_{\rm WW}&=&-(-J/4+J_0), 
\nonumber\\
\lambda_{\rm BW}&=&-(+J/4+J_0). 
\nonumber
\end{eqnarray}
The overall minus signs are introduced because 
the probability to find a configuration with total adhesion strength $\Lambda_k$ 
is proportional to $\exp(\Lambda_k/m)$ in the case of the cell-sorting model, 
though the probability to find a configuration with total energy $E_k$ 
is proportional to $\exp(-E_k/k_{\rm B}T)$ in the case of the Ising model. 
It is easy to check that the contributions from $J_0$ cancel 
in the calculations of expectation values.  
The Hamiltonian is 
$H=-J\sum_{\langle ij\rangle}s_i^zs_j^z+\frac{1}{2}zNJ_0$, 
where $N$ is the total number of sites 
and $z$ is the number of interacting pairs $\langle ij\rangle$ from each site $i$,  
e.g. $z=2$ in the one-dimensional linear chain 
and $z=4$ in the two-dimensional square lattice. 
The parameters in the cell-sorting model and those in the Ising model 
are related by $A=J$ and $m=k_{\rm B}T$, and hence $A/m=J/k_{\rm B}T$. 
The probability of re-arrangement increase as the temperature increase. 
The correspondence is consistent with the facts that 
$J$ is the coupling constant which determines the unit of energy scale in the Ising model, 
and $A$ is the the difference adhesion which determines the unit of the total adhesion 
in the cell-sorting problem. 

As a model of magnetism, $A/m=-2$ indicates that the Ising model is an antiferromagnet. 
The checkerboard-like pattern is that called the N\'{e}el order. 
The case with $A/m=0$ corresponds to a paramagnet, 
where there is no interaction between spins, and each spin flips independently. 
The cases with positive $A/m$ correspond to ferromagnets, 
where the spin of the nearest-neighbors tend to become parallel and tend to form clusters. 
In the case studied by Mochizuki et al., 
the number of the black cells and the number of the white cells are fixed to be the same. 

The two-dimensional Ising model has a critical point $T_c$ 
at a non-zero and finite value of temperature.  
Below $T_c$, clusters of spin up states appear, 
which corresponds to the existence of cell-sorting phenomena, 
whereas above $T_c$ there is no cell-sorting. 
In the case of the one-dimensional Ising model, the critical temperature is equal to zero. 

The quantities studied in the cell-sorting problem 
correspond to quantities in the Ising model. 
Let us assume the periodic boundary condition. 
Following the notations in Mochizuki et al., 
let 
$\rho_{\rm B}$ be the fraction of the black cells and 
$\rho_{\rm W}$ be the fraction of the white cells, 
then $\rho_{\rm B}+\rho_{\rm W}=1$. 
Let 
$q_{\rm BB}$ be the fraction of black cells 
in the neighborhood of a randomly chosen black cell, 
$q_{\rm WW}$ the fraction of white cells 
in the neighborhood of a randomly chosen white cell, 
$q_{\rm WB}$ the fraction of white cells 
in the neighborhood of a randomly chosen black cell, and  
$q_{\rm BW}$ the fraction of black cells 
in the neighborhood of a randomly chosen white cell. 
Then $q_{\rm BB}+q_{\rm WB}=1$ and $q_{\rm WW}+q_{\rm BW}=1$. 
The two-body correlation function in the Ising model is the expectation value 
of the product of eigenvalues $\pm 1/2$ of corresponding two sites. 
Thus, by definition, the nearest-neighbor correlation function is written 
using the probabilities as
\begin{eqnarray}
\langle s_{i}^zs_{j}^z\rangle_{\rm Ising}
&=&\frac{1}{2}\frac{1}{2}\rho_{\rm B}\: q_{\rm BB}
+\frac{1}{2}\frac{-1}{2}\rho_{\rm B}\: q_{\rm WB}
+\frac{-1}{2}\frac{1}{2}\rho_{\rm W}\: q_{\rm BW}
+\frac{-1}{2}\frac{-1}{2}\rho_{\rm W}\: q_{\rm WW}
\nonumber\\
&=&\frac{1}{4}\rho_{\rm B}\: (2q_{\rm BB}-1)+\frac{1}{4}\rho_{\rm W}\: (2q_{\rm WW}-1)
\nonumber\\
&=&\frac{1}{4}(2\rho_{\rm B}\: q_{\rm BB}+2\rho_{\rm W}\: q_{\rm WW}-1).  
\label{prob2}
\end{eqnarray}
The magnetization $\langle s_{i}^z\rangle_{\rm Ising}$ is the expectation value 
of the eigenvalue associated to each site, 
and obviously is equal to $\frac{1}{2}\rho_{\rm B}+(-\frac{1}{2})\rho_{\rm W}$. 
However, it can also be written in terms of the probabilities 
$\rho_{\rm B}\: q_{\rm BB}$ and $\rho_{\rm W}\: q_{\rm WW}$.
The probability that the two cells in a randomly chosen nearest-neighbor pair are both black 
is $\rho_{\rm B}\: q_{\rm BB}$, 
the probability that one of the two cells is black and the other cell is white 
is $\rho_{\rm B}\: q_{\rm WB}+\rho_{\rm W}\: q_{\rm BW}$, and  
the probability that the two cells are both white 
is $\rho_{\rm W}\: q_{\rm WW}$.  
Considering the average magnetization per pair, we obtain 
\begin{eqnarray}
\langle s_i^z\rangle_{\rm Ising}
&=&\frac{1}{2}
[(\frac{1}{2}+\frac{1}{2})\rho_{\rm B}\: q_{\rm BB}
+(\frac{1}{2}-\frac{1}{2})(\rho_{\rm B}\: q_{\rm WB}+\rho_{\rm W}\: q_{\rm BW})\nonumber\\
&&+(-\frac{1}{2}-\frac{1}{2})\rho_{\rm W}\: q_{\rm WW}]
\nonumber\\
&=&\frac{1}{2}(\rho_{\rm B}\: q_{\rm BB}-\rho_{\rm W}\: q_{\rm WW})
\label{prob1}
\end{eqnarray}
From (\ref{prob2}) and (\ref{prob1}), 
$\rho_{\rm B}q_{\rm BB}$  and $\rho_{\rm W}q_{\rm WW}$ 
are expressed by the correlation functions in the Ising model as 
\begin{eqnarray}
\rho_{\rm B}\: q_{\rm BB}
&=&
\langle s_i^zs_j^z\rangle_{\rm Ising}+\frac{1}{4}+\langle s_i^z\rangle_{\rm Ising},
\nonumber\\
\rho_{\rm W}\: q_{\rm WW}
&=&
\langle s_i^zs_j^z\rangle_{\rm Ising}+\frac{1}{4}-\langle s_i^z\rangle_{\rm Ising},
\label{corr2}
\end{eqnarray}
This argument can be generalized to obtain probabilities for other types of spin pairs.  
When the site $i$ and $j$ locate 
in the next-nearest-neighbors (or in other specific locations) each other, 
$q_{\rm BB}$ etc. should be redefined as the fraction 
to find cells  of corresponding colors 
in the next-nearest-neighbors (or in the other specific locations).  
Then  (\ref{corr2}) is still valid.   
With the open boundary,  
(\ref{corr2}) is also valid except boundary terms, 
which vanish when one assume the periodic boundary condition. 

In the one-dimensional Ising model, 
there is no spontaneous symmetry breaking, 
and hence $\langle s_i^z\rangle_{\rm Ising}=0$, thus $\rho_{\rm B}=1/2$.  
The nearest-neighbor correlation function of this case 
is immediately calculated by the transfer matrix method and one obtains 
$q_{\rm BB}=2(\tanh(J/4kT)/4+1/4+0)=(\tanh(A/4m)+1)/2$, 
which is consistent with the result obtained for the cell-sorting model.

\section{Equivalences between biological systems}
The cell-sorting model with adhesion strength 
is equivalent to the Ising model in each dimension. 
The two-dimensional Ising model on the square lattice 
is equivalent to the one-dimensional XY model with an external magnetic field, 
and it is also equivalent to the one-dimensional transverse Ising model. 
The XY model and the transverse Ising model 
are equivalent to systems with two-body flip, or independent one-body flip, 
with the energy estimated as the product of eigenvalues of two states. 
Next I consider a stochastic process that corresponds to the XY model. 

Let us consider identical particles on the one-dimensional lattice. 
Each particle locate on a lattice point $x$, where $x$ is integer. 
The particles move right or left in each step 
with the rate $p_{\rm R}$ and $p_{\rm L}$, respectively. 
The particles interact with the hard-core interactions: 
they cannot move into the space already occupied by other particles. 
Let us introduce the following notation 
\begin{eqnarray}
|-\cdots, +, -, \cdots+\rangle= |-\rangle_1\cdots|+\rangle_j|-\rangle_{j+1}\cdots|+\rangle_m, \hspace{0.6cm}{\rm etc.}
\nonumber
\end{eqnarray}
Operators $s_j^\mp$ and $s_{j+1}^\pm$ work nontrivially 
only on the $j$-th and $j+1$-th sites as
\begin{eqnarray}
s_j^-s_{j+1}^+|\cdots, +, -, \cdots\rangle&=&|\cdots, -, +, \cdots\rangle,\nonumber\\
s_j^+s_{j+1}^-|\cdots, -, +, \cdots\rangle&=&|\cdots, +, -, \cdots\rangle,
\label{hop}
\end{eqnarray}
When we regard $+$ as a particle and $-$ as an empty site, 
operations (\ref{hop}) are the mappings 
in which a particle moves to one of its nearest neighbors. 
The matrix representations of the operators are  
\begin{eqnarray}
s_j^-s_{j+1}^+=
\left[
\begin{array}{cccc}
0&0&0&0\\
0&0&0&0\\
0&0&0&0\\
0&0&1&0
\end{array}
\right]_{jj+1},
\hspace{0.3cm}
s_j^+s_{j+1}^-=
\left[
\begin{array}{cccc}
0&0&0&0\\
0&0&0&0\\
0&0&0&1\\
0&0&0&0
\end{array}
\right]_{jj+1},
\label{s+s-}
\end{eqnarray}
where the basis set used to represent the matrix is 
\begin{eqnarray}
\{ |+\rangle_j|+\rangle_{j+1}, |-\rangle_j|-\rangle_{j+1}, 
|+\rangle_j|-\rangle_{j+1}, |-\rangle_j|+\rangle_{j+1}, \}.  
\nonumber
\end{eqnarray}
The matrix $[\;]_{jj+1}$ operates on the site $j$ and $j+1$, 
and operates as an identity on the other sites. 
Then the contributions from the random hoppings are written as    
\begin{eqnarray}
H&=&\sum_{j=1}^N[p_{\rm R}s_j^-s_{j+1}^++p_{\rm L}s_j^+s_{j+1}^-] 
\label{pXY}\\
&=&\sum_{j=1}^N
\left[
\begin{array}{cccc}
0&0&0&0\\
0&0&0&0\\
0&0&0&p_{\rm L}\\
0&0&p_{\rm R}&0
\end{array}
\right]_{jj+1}. 
\label{H}
\end{eqnarray}
Let $\{x_1, x_2, \ldots, x_{m_{\rm p}}\}$ be a configuration of the particles, 
in which the sites $x_1, x_2, \ldots, x_{m_{\rm p}}$ are occupied 
and the other sites are empty. 
Let $\{x_i\}_k \;\;(k=1, \ldots, 2^N)$ be the possible $2^N$ configurations, 
where each $k$ denotes one configuration. 
Let  $P(\{x_i\}_k ; n)$ be the probability 
that the configuration of the particles is $\{x_i\}_k$ after the $n$-th step. 
Let us introduce the probability vector ${\bf P}_n$, 
in which the $k$-th element of ${\bf P}_n$ is $({\bf P}_n)_k=P(\{x_i\}_k ; n)$. 
Then the random hoppings are generated by ${\bf P}_{n+1}=H{\bf P}_n$.  
This process conserves the number of particles. 
In the case of $p_{\rm R}=p_{\rm L}=1/2$, 
the operator (\ref{pXY}) is reduced to the Hamiltonian of the XY-model (\ref{XY}) 
with $\gamma=0$ and $h=0$. 

When one introduce nearest-neighbor pair creation and annihilation processes 
with the probability $p_{\rm U}$ and $p_{\rm D}$, respectively, 
and also multiply the weights $-h/2$ and $h/2$ 
for the nearest-neighbor pair particles and pair of the empty sites, respectively, 
$H$ becomes  
\begin{eqnarray}
H=\sum_{j=1}^N
\left[
\begin{array}{cccc}
-h/2&p_{\rm U}&0&0\\
p_{\rm D}&h/2&0&0\\
0&0&0&p_{\rm L}\\
0&0&p_{\rm R}&0
\end{array}
\right]_{jj+1}.
\label{H2}
\end{eqnarray}
This operator (\ref{H2}) is written by the spin operators as 
\begin{eqnarray}
H=\sum_{j=1}^N[p_{\rm R}s_j^-s_{j+1}^++p_{\rm L}s_j^+s_{j+1}^-
                                  +p_{\rm U}s_j^+s_{j+1}^++p_{\rm D}s_j^-s_{j+1}^-
                                  -\frac{h}{2}(s_j^z+s_{j+1}^z)].
\label{pXY2}
\end{eqnarray}
When $p_{\rm R}=p_{\rm L}=1/2$ and $p_{\rm U}=p_{\rm D}=\gamma/2$,  
(\ref{pXY2}) is reduced to the Hamiltonian 
of the anisotropic XY-model (\ref{XY}) with an external field $h$. 

The first $2\times 2$ block element in (\ref{H2}) 
represents the following processes.  
Two particles meet and annihilate with the rate $p_{\rm D}$, 
a nearest-neighbor pair of particles are created with the rate $p_{\rm U}$.  
The weight of each state varies spontaneously: 
multiplied by the factor 
which is the sum of $-h/2$ and $h/2$ 
associated to each nearest-neighbor pair of particles 
and each nearest-neighbor pair of empty sites, respectively. 

Let $p_{\rm U}=p_{\rm D}=\gamma/2$, 
and assume $\gamma>0$. 
All the matrix elements in the first block in (\ref{H2}) can become non-negative 
by a rotation as  
\begin{eqnarray}
R^{-1}
\frac{1}{2}
\left[
\begin{array}{cc}
-h&\gamma\\
\gamma&h
\end{array}
\right]_{jj+1}
R
=
\lambda
\left[
\begin{array}{cc}
0&1\\
1&0
\end{array}
\right]_{jj+1}, 
\nonumber
\end{eqnarray}
where $\lambda=\sqrt{\gamma^2+h^2}/2$ and 
\begin{eqnarray}
R=
\left[
\begin{array}{cc}
\cos\theta&-\sin\theta\\
\sin\theta&\cos\theta
\end{array}
\right]_{jj+1},
\hspace{0.2cm}
\tan\theta=
\frac{h-\gamma+\sqrt{\gamma^2+h^2}}{h+\gamma+\sqrt{\gamma^2+h^2}}, 
\hspace{0.2cm}
|\theta|<\frac{\pi}{4}.
\nonumber
\end{eqnarray}
Thus the pair creation and annihilation processes 
with the spontaneous change $-h/2$ and $h/2$ 
can also be regarded as a simple reflection multiplied by the factor $\lambda$, 
in the above two rotated basis.   

Therefore the Hamiltonian of the XY-model 
can be regarded as the generator of the following two processes, 
1) random walk process with hard-core interactions, and 
2) pair creations and pair annihilations 
with spontaneous change of weights 
which are determined from the number of pair particles and "pair vacuums".  
The second process 2) is also regarded as the rotated reflections 
with the rate $\sqrt{\gamma^2+h^2}/2$. 

In the relation (\ref{corr1}), 
the expectation values $\langle\:\rangle_{\rm 1D\:XY}$ 
are estimated in the ground state $|\phi_0\rangle$ of the XY model  
with the coupling constant $J>0$, and thus $-J<0$. 
The state $|\phi_0\rangle$ 
is the eigenstate corresponding to the maximum eigenvalue of (\ref{pXY2})
because the sign of the coupling is now $-J=1>0$. 
When we consider the matrix $H+cI$ 
where $I$ is the unit matrix and $c>0$ is sufficiently large,  
all the matrix elements of $H+cI$ are non-negative, 
and the eigenstates of $H+cI$ are simultaneously the eigenstates of $H$. 
From the Perron-Frobenius theorem, 
it is derived that all the coefficients in $|\phi_0\rangle$ are non-negative. 

Let $\lambda_0$ be the maximum eigenvalue of $H$. 
The maximum eigenvalue of the matrix ${\bar H}=H/\lambda_0$ is equal to $1$. 
Let $\Lambda_i$ be an eigenvalue of ${\bar H}$ and 
$|\phi_i\rangle$ be an eigenstate  corresponding to $\Lambda_i$: 
${\bar H}|\phi_i\rangle=\Lambda_i|\phi_i\rangle$.  
Then all the eigenvalues satisfy $|\Lambda_i|\leq |\Lambda_0|=1$. 
Therefore the state $|\phi_0\rangle$ survives 
in the limit where the operator ${\bar H}$ is iteratively applied:  
${\bar H}^n|\phi_0\rangle=|\phi_0\rangle$. 

In the one-dimensional XY model, there exist a critical field $h_c$, 
which corresponds to the critical temperature $T_c$ of the two-dimensional Ising model.   
The ground state is unique for $h>h_c$ and two-fold degenerate for $h\leq h_c$ 
in the thermodynamic limit $N\to\infty$. 
In the case of the two-dimensional Ising model, 
the eigenstate of the transfer matrix corresponding to the maximum eigenvalue  
is unique for $T>T_c$ and two-fold degenerate for $T\leq T_c$,  
and there exists a spontaneous symmetry breaking 
and the magnetization becomes non-zero below $T_c$. 
This implies that the global minimum of the free energy is realized 
with $\rho_B=1/2$ for $T>T_c$, and with $\rho_B\neq 1/2$ for $T\leq T_c$. 
The corresponding cell-sorting model below $T_c$ is, therefore, 
a generalized model which includes the effect of gathering of one kind of cells from outside.  

If the process represented by ${\bar H}$ is fully stochastic, 
the matrix ${\bar H}$ should satisfy the conservation of probability: 
$({\bar H})_{ij}\geq 0$ and $\sum_i({\bar H})_{ij}=1$.  
The matrix ${\bar H}$ does not satisfy this condition. 
However, the state $|\phi_0\rangle$ is a steady state 
and the conservation of probability is satisfied in $|\phi_0\rangle$. 
Let 
\begin{eqnarray}
|\phi_0\rangle
=\sum_{\{x_i\}}P_0(x_1, x_2, \ldots, x_{m_{\rm p}} ; n)|x_1, x_2, \ldots, x_{m_{\rm p}}\rangle,
\nonumber
\end{eqnarray}
where $|x_1, x_2, \ldots, x_{m_{\rm p}}\rangle$ is 
the direct product of the states $ |+\rangle_j \;\;(j=x_1, x_2, \ldots, x_{m_{\rm p}})$ 
and $ |-\rangle_j \;\;(j\neq x_1, x_2, \ldots, x_{m_{\rm p}})$, 
i.e. the state where the sites $x_1, \ldots, x_{m_{\rm p}}$ are occupied 
and the other sites are empty.    
The coefficient $P_0(x_1, x_2, \ldots, x_{m_{\rm p}} ; n)$ is the probability 
that the system is in the state $|x_1, x_2, \ldots, x_{m_{\rm p}}\rangle$. 
The probability distribution $\{P_0(x_1, x_2, \ldots, x_{m_{\rm p}} ; n)\}$ 
is invariant under the operations of ${\bar H}$ 
and satisfies the conservation of probability.  
This means that ${\bar H}$ is stochastic in the subspace corresponding to the maximum eigenvalue. 

In the case of the cell-sorting problem 
on the square grid, 
from (\ref{corr1}), (\ref{corr2}), 
and from the relations 
$\langle s_i^z\rangle_{\rm 2D Ising}=(1/2)\rho_{\rm B}+(-1/2)\rho_{\rm W}$, 
$\rho_{\rm B}+\rho_{\rm W}=1$, 
and $\cosh 2K_1^*=1/\gamma$, 
one obtains 
\begin{eqnarray}
\rho_{\rm B}q_{\rm BB}
&=&\langle s_{ij}^zs_{ik}^z\rangle_{\rm 2D Ising}+\frac{1}{4}
+\langle s_i^z\rangle_{\rm 2D Ising}
\nonumber\\
&=&\cosh^2K_1^*\langle s^x_js^x_k\rangle_{\rm 1D \:XY}
-\sinh^2K_1^*\langle s^y_js^y_k\rangle_{\rm 1D \:XY}+\frac{1}{4}
+(\rho_{\rm B}-\frac{1}{2})
\nonumber\\
&=&\rho_{\rm B}
+\frac{1}{4}[\langle s_j^+s_k^-+s_j^-s_k^+\rangle_{\rm 1D \:XY}
+\frac{1}{\gamma}\langle s_j^+s_k^++s_j^-s_k^-\rangle_{\rm 1D \:XY}-1]
\nonumber\\
\label{corr3}
\end{eqnarray}
The probability $q_{\rm BB}$ in (\ref{corr3}) is that 
for the two cells locating $(i, j)$ and $(i, k)$. 
The parameter $\gamma$ is specified from the restriction 
$K_1=K_2=\beta J=A/m$. 
The expectation values are calculated in principle 
from $\langle\:\cdot\:\rangle_{\rm 1D \:XY}=\langle\phi_0|\cdot|\phi_0\rangle$, 
the expansion (\ref{innerproduct}), 
and the normalization conditions 
$\langle+|+\rangle=1$, $\langle-|-\rangle=1$, 
$\langle+|-\rangle=0$ and $\langle-|+\rangle=0$ for each site $i$. 
The magnetization $\langle s_i^z\rangle_{\rm 2D Ising}$ below $T_c$ 
have been calculated by Yang (1952). 
The correlation functions $\langle s^l_js^l_k\rangle_{\rm 1D \:XY} \:\:(l=x, y)$ 
have been analytically calculated 
in the context of the one-dimensional XY model 
(McCoy et al., 1971; Tonegawa, 1981), 
and $\langle s^z_{00}s^z_{jk}\rangle_{\rm 2D \:Ising}$ 
have also been calculated in the context of the equivalent two-dimensional Ising model 
(Wu et al. 1976; McCoy et al. 1977), 
both for arbitraly $j$ and $k$.  
In conclusion, the adhesion probability in the two-dimensional cell-sorting problem 
is analytically expressed by the expectation values in the steady state  
of the one-dimensional generalized random walk problem.

\section{Correspondences of stochastic processes with lattice spin models}
The matrix $H$ (and ${\bar H}$) corresponding to the XY model 
is not itself fully stochastic for all the possible states of the particles. 
In this section, 
first I formulate the most general form of the stochastic matrix 
for the one-dimensional random walk process with creations and annihilations,   
and next consider three examples to show the close relation 
between biological systems and lattice spin systems, 
and finally prove that  
the stochastic processes in biological systems 
can always be written in terms of the spin Hamiltonians. 

Let us consider the operator 
\begin{eqnarray}
H=\sum_{j=1}^N
\left[
\begin{array}{cccc}
w_3&p_{\rm U}&0&0\\
p_{\rm D}&w_4&0&0\\
0&0&w_1&p_{\rm L}\\
0&0&p_{\rm R}&w_2
\end{array}
\right]_{jj+1}.
\label{Hgen}
\end{eqnarray}
The element $p_{\rm D}$ in the matrix $[\hspace{0.1cm}]_{jj+1}$ 
is the rate  of the pair annihilation 
$|+\rangle_j|+\rangle_{j+1}\mapsto |-\rangle_j|-\rangle_{j+1}$.  
The element $w_3$ is the rate of the pair to be invariant 
$|+\rangle_j|+\rangle_{j+1}\mapsto |+\rangle_j|+\rangle_{j+1}$,  
and the other elements are defined as the rates of the following transitions:  
\begin{eqnarray}
&&p_{\rm U}:|-\rangle_j|-\rangle_{j+1}\mapsto |+\rangle_j|+\rangle_{j+1},\hspace{0.6cm}
w_4:|-\rangle_j|-\rangle_{j+1}\mapsto |-\rangle_j|-\rangle_{j+1} 
\nonumber\\
&&p_{\rm R}:|+\rangle_j|-\rangle_{j+1}\mapsto |-\rangle_j|+\rangle_{j+1},\hspace{0.6cm}
w_1:|+\rangle_j|-\rangle_{j+1}\mapsto |+\rangle_j|-\rangle_{j+1}
\nonumber\\
&&p_{\rm L}:|-\rangle_j|+\rangle_{j+1}\mapsto |+\rangle_j|-\rangle_{j+1},\hspace{0.6cm}
w_2:|-\rangle_j|+\rangle_{j+1}\mapsto |-\rangle_j|+\rangle_{j+1}.
\nonumber
\end{eqnarray}
First let us assume the periodic boundary condition, 
and consider the case $w_i=0\:(i=1, 2, 3, 4)$. 
Starting from the state 
$|+\rangle_1|+\rangle_2\cdots |+\rangle_N=|++\cdots +\rangle$, 
$N$ kind of pair annihilations 
which result in the states with successive two minus, 
$|--++\cdots +\rangle$, $|+--+\cdots +\rangle$, etc., are possible. 
Thus the conservation of probability is satisfied with the condition $Np_{\rm D}=1$. 
Starting from other states such as 
$|+-++\cdots +\rangle$, $|+--+\cdots +\rangle$, $|+-+-\cdots +\rangle$, etc., 
it is straightforward to convince that 
two $p_{\rm D}$'s (or two $p_{\rm U}$'s) 
are always replaced by a pair formed by $p_{\rm R}$ and $p_{\rm L}$ 
in the calculations of the total probability, 
and thus the conservation of probability is satisfied with  
$2p_{\rm D}=p_{\rm R}+p_{\rm L}$, 
$2p_{\rm U}=p_{\rm R}+p_{\rm L}$.  
Similarly in the case of $p_{\rm D}$, 
the condition for $p_{\rm U}$ is $Np_{\rm U}=1$. 
Thus we obtain $p_{\rm D}=p_{\rm U}=1/N$, and $p_{\rm R}+p_{\rm L}=2/N$. 
Let us introduce the rates $w_i\:(i=1, 2, 3, 4)$, 
then $p_{\rm D}$, $p_{\rm U}$, $p_{\rm R}$, and $p_{\rm L}$ 
should be replaced by 
$p_{\rm D}+w_3$, $p_{\rm U}+w_4$, $p_{\rm R}+w_1$, and $p_{\rm L}+w_2$, 
respectively, in the calculations of the total probability. 
Therefore the conservation of probability is satisfied iff 
\begin{eqnarray}
p_{\rm D}+w_3=p_{\rm U}+w_4=\frac{1}{N},
\label{condstoch1}\\
(p_{\rm R}+w_1)+(p_{\rm L}+w_2)=\frac{2}{N}.
\label{condstoch2}
\end{eqnarray}
When we assume the open boundary condition, 
$s_{N+1}^z\neq s_1^z$, then the number of spin pairs $(j, j+1)$ is $N$, 
and the condition $(\ref{condstoch1})$ remains true 
though the number of sites now being $N+1$. 
The condition $(\ref{condstoch2})$ should be satisfied for $2\leq j\leq N-1$.    
At $(j, j+1)=(1, 2)$ and $(N, N+1)$, the probability is conserved iff 
\begin{eqnarray}
p_{\rm R}+w_1=p_{\rm L}+w_2=\frac{1}{N}.
\label{condstoch3}
\end{eqnarray}
The condition $(\ref{condstoch2})$ is satisfied when we assume $(\ref{condstoch3})$. 
Thus $(\ref{condstoch1})$ and $(\ref{condstoch3})$ 
are the conditions in the case of the lattice with open boundary. 
These are the general form of the one-dimensional stochastic matrix 
for random walk processes, with pair creations and pair annihilations. 

The matrix (\ref{Hgen}) is written by the spin operators as 
\begin{eqnarray}
H=\sum_{j=1}^N[
 p_{\rm R}s_j^-s_{j+1}^++p_{\rm L}s_j^+s_{j+1}^-
+p_{\rm U}s_j^+s_{j+1}^++p_{\rm D}s_j^-s_{j+1}^-.\nonumber\\
+\Delta s_j^zs_{j+1}^z-\frac{h}{2}(s_j^z+s_{j+1}^z)
+c_0I+\frac{c}{2}(s_j^z-s_{j+1}^z)],
\label{pXY3}
\end{eqnarray}
where $I$ is the identity operator and 
\begin{eqnarray}
\Delta&=&w_3+w_4-w_1-w_2,\hspace{0.6cm}
h=w_4-w_3\nonumber\\
c_0&=&\frac{1}{4}(w_3+w_4+w_1+w_2),\hspace{0.6cm}
c=w_1-w_2.
\label{pXY2c}
\end{eqnarray}

When one assume the periodic boundary condition $s_{N+1}^z=s_1^z$, 
the last term in (\ref{pXY3}) vanishes: $\sum_{j=1}^N(s_j^z-s_{j+1}^z)=0$, 
and thus $H$ is independent of $c$. 
It is easy to check that $w_3$, $w_4$ and $w_1+w_2$ do not depend on $c$, 
and hence the process remains stochastic, 
i.e. satisfy (\ref{condstoch1}) and (\ref{condstoch2}), with $c$ being a free parameter.  
This is a result which comes from the translational invariance of the system. 

Next let us consider one-body creations and annihilations of the particles.   
Let us introduce an operator 
\begin{eqnarray}
n_j=\frac{1}{2}+s_j^z. 
\label{nj} 
\end{eqnarray}
In our notation, the eigenstate of $s_j^z$ with the eigenvalue $+1/2$ (the eigenvalue $-1/2$) 
corresponds to the state with a particle (without a particle) at site $j$. 
The operator $n_j$ takes the value $1$ or $0$ 
when the site $j$ is occupied by a particle or the site $j$ is empty, 
respectively. 
Thus $n_j$ is called the number operator. 
Let us consider the case where a particle is added at site $j$. 
The new state with the added particle is created by $s_j^+$, 
and the old state without the particle is removed by $-(1-n_j)$: 
\begin{eqnarray}
\left[
\begin{array}{cc}
0&1\\
0&0
\end{array}
\right]_j
-
\left[
\begin{array}{cc}
0&0\\
0&1
\end{array}
\right]_j
=s_j^+-(1-n_j). 
\nonumber
\end{eqnarray}
(Note that these operators work non-trivially only in the subspace where $n_j=0$, 
states in which the site $j$ is empty, 
and otherwise they  work as the zero operator.)  
Thus the term $\Delta H_j^+=\alpha_j[s_j^+-(1-n_j)]$ should be added 
to the Hamiltonian $H$ in (\ref{pXY3}), 
where $\alpha_j$ is the rate of creation at site $j$. 
Similarly, if a particle at site $j$ is annihilated  with the rate $\beta_j$, 
the term $\Delta H_j^-=\beta_j[s_j^--n_j]$ should be added to (\ref{pXY3}). 
(These operators work only in the subspace where $n_j=1$, 
states with a particle at site $j$.) 
Because $s_j^\pm=s_j^x\pm is_j^y$, 
these terms are the transverse and parallel magnetic fields applied to the site $j$. 
Thus we find that external fields induce 
the spontaneous creations and annihilations (attachments and displacements) 
of the particles.  
\newline

\noindent
{\bf Random walk with hard-core interaction:} 
From (\ref{H}) and (\ref{condstoch1}) -(\ref{pXY2c}),  
the fully stochastic random walk process with hard-core interactions 
is written as 
\begin{eqnarray}
H_{\rm RW}=\frac{1}{N}\sum_{j=1}^N
\left[
\begin{array}{cccc}
1&0&0&0\\
0&1&0&0\\
0&0&0&\eta_{\rm L}\\
0&0&\eta_{\rm R}&0
\end{array}
\right]_{jj+1},
\label{RW1}
\end{eqnarray}
where $\eta_{\rm R}/N=p_{\rm R}$, $\eta_{\rm L}/N=p_{\rm L}$ 
and $\eta_{\rm R}+\eta_{\rm L}=2$. 
This operator is written by the spin operators as  
\begin{eqnarray}
H_{\rm RW}=\frac{2}{N}
\sum_{j=1}^N[\frac{1}{2}(\eta_{\rm R}s_j^-s_{j+1}^++\eta_{\rm L}s_j^+s_{j+1}^-)
                                                  +s_j^zs_{j+1}^z +\frac{1}{4}I]. 
\label{RW2}
\end{eqnarray}
When $\eta_{\rm R}=\eta_{\rm L}=1$, 
the operator (\ref{RW2}) is the Hamiltonian of the XXZ model (\ref{XXZ}) with $\Delta=1$, 
i.e. the Hamiltonian of the Heisenberg model.  
This  case is a kind of the hard-core boson system,
which is originally proposed as a model for helium superfluidity 
(Matsubara and Matsuda, 1956). 

The anisotropy $\eta_{\rm R}\neq \eta_{\rm L}$ 
can be removed by an unitary transformation 
(Henkel and Sch\"{u}tz 1994, 
in which the eigenstate with the eigenvalue $-1/2$ 
is regarded as the state with a prticle) 
provided $\eta_{\rm R}\eta_{\rm L}\neq 0$.  
Let 
\begin{eqnarray}
V=\exp[(\log q)\sum_{j=1}^N jn_j], 
\label{transV} 
\end{eqnarray}
where $n_j$ is given by (\ref{nj}). 
The commutation relation 
$[n_j, s_j^\pm]=\pm s_j^\pm$ and the expansion 
\begin{eqnarray}
e^L Ae^{-L}
=A+\frac{1}{1!}[L, A]+\frac{1}{2!}[L,[L, A]]+\frac{1}{3!}[L,[L,[L, A]]]+\cdots, 
\nonumber
\end{eqnarray}
yield $Vs_j^\pm V^{-1}=q^{\pm j}s_j^\pm$ and $Vs_j^z V^{-1}=s_j^z$. 
Let $q=\sqrt{\eta_{\rm L}/\eta_{\rm R}}$ and one obtains 
\begin{eqnarray}
VH_{\rm RW}V^{-1}
&=&V[
\frac{2}{N}
\sum_{j=1}^N[
\frac{\sqrt{\eta_{\rm R}\eta_{\rm L}}}{2}(q^{-1} s_j^-s_{j+1}^++qs_j^+s_{j+1}^-)
                                                 +s_j^zs_{j+1}^z +\frac{1}{4}I]
V^{-1}
\nonumber\\
&=&\frac{2}{N}\sqrt{\eta_{\rm R}\eta_{\rm L}}
\sum_{j=1}^N[\frac{1}{2}(s_j^-s_{j+1}^++s_j^+s_{j+1}^-)
                                                 +(s_j^zs_{j+1}^z +\frac{1}{4}I)/\sqrt{\eta_{\rm R}\eta_{\rm L}}].
\nonumber
\end{eqnarray}
Hence it is derived that 
the anisotropy of the hopping rates in the random walk process 
can be handled as the anisotropy 
$\Delta=1/\sqrt{\eta_{\rm R}\eta_{\rm L}}$ 
of the quantum coupling  in an uniform spin chain. 
\newline

\noindent
{\bf Ribosome moving on mRNA:} 
A ribosome is a large, complex molecule 
that  synthesizes a protein molecule 
using the genetic message coded on mRNA as the template. 
RNA comprises four kinds of nucleotides, 
and a triplet of nucleotides constitutes a codon. 
Each possible type of codon corresponds to one species of amino acid: 
61 kinds of codon lead to 20 species of amino acid, 
whereas three special codons indicate termination of translation. 
The information enclosed in the codon sequence 
is translated by the ribosome 
into the amino acid sequence of the encoded proteins. 

A ribosome binds to an mRNA 
and begins to synthesize the protein by adding an amino acid 
(this is referred to as initiation). 
After a biochemical reaction for the elongation of the protein, 
the ribosomes moves forward on the track by one codon 
(i.e. one elongation has occured). 
Finally, the ribosome reaches the termination codon 
and leaves the mRNA, releasing the protein (this is referred to as termination). 

A number of ribosomes can be simultaneously attached to one mRNA template. 
A ribosome can move forward on the track 
provided that the next codon is not captured by another ribosome, 
i.e. ribosomes are moving on the mRNA template 
interacting via hard-core interactions. 
A ribosome is often treated as a molecular motor, 
and this collective movement process along the mRNA chain shows a correspondence 
with a one-dimensional driven lattice gas, or with vehicular traffic on a road.

MacDonald et al. introduced a stochastic process now known as the asymmetric simple exclusion process (ASEP), as a model for the movement of ribosomes on a mRNA  
(MacDonald et al. 1968; MacDonald and Gibbs 1969). 
The model was first introduced in the biophysical literature, 
and later studied from a purely theoretical viewpoint 
(see for example  Derrida 1998). 

The actual movement of a ribosome is closely coupled to its internal mechanochemical processes to synthesize a protein. 
Accounting for these processes, the ASEP has been generalized 
to have seven  
(Basu and Chowdhury 2007), 
five 
(Garai et al. 2009), 
or two 
(Ciandrini et al.)
distinct biochemical states in each cycle. 
The ribosome movement is also characterized by a pause and translocation, 
which defines the time of its residing at a corresponding codon. 
The ASEP has been generalized to have one or more slow codon bottlenecks 
(Kolomeisky 1998; Chou and Lakatos 2004; Dong et al. 2007a; Dong et al. 2007b). 
The ASEP with different hopping rates associated with each site has been considered 
(Shaw et al. 2003; Shaw et al. 2004; Romano et al. 2009).  
A ribosome recycling mechanism has been introduced 
(Chou 2003), 
in which a part of the ribosome detaches at the termination site 
and part of them diffuses back to the initiation site.  
The ASEP comprising open boundaries with random particle attachments and detachments 
was also introduced (Parmeggiani et al. 2003, Pierobon et al. 2006). 
A stochastic model with a secondary structure of mRNA was introduced  
by von Heijine et al. (1977).  

Now, let us introduce the ASEP, 
which is a random walk process with specified moving rates 
and with a continuous time variable. 
Let us consider an one-dimensional lattice. 
Each site $j$ ($j\in{\bf Z}$) is empty or occupied by a particle 
which corresponds to a ribosome.  
Each particle stochastically move forward or backward 
interacting via the hard-core interactions i.e. 
a particle at site $j$ moves to the site $j+1$ with the rate $p_{\rm R}$ 
provided that the site $j+1$ is empty, 
and a particle at site $j+1$ moves to the site $j$ with the rate $p_{\rm L}$ 
provided that the site $j$ is empty. 

Let us first assume the periodic boundary condition  
and consider the stochastic matrix  (\ref{Hgen}) 
with $p_{\rm U}=p_{\rm D}=0$, and thus from (\ref{condstoch1})  $w_3=w_4=1/N$. 
Because $H$ is independent of $c$, 
the rates $w_1$ and $w_2$ can be taken as 
\begin{eqnarray}
w_1=\frac{1}{N}(1-\delta)-p_{\rm R},\hspace{0.3cm}
w_2=\frac{1}{N}(1+\delta)-p_{\rm L},\hspace{0.6cm}
(c=p_{\rm L}-p_{\rm R}-\frac{2}{N}).
\nonumber
\end{eqnarray}
These rates satisfy (\ref{condstoch1}) and (\ref{condstoch2}). 
The corresponding stochastic matrix is 
\begin{eqnarray}
H_{\rm ASEP}&=&\frac{1}{N}\sum_{j=1}^N
\left[
\begin{array}{cccc}
1&0&0&0\\
0&1&0&0\\
0&0&(1-\delta)-\eta_{\rm R}&\eta_{\rm L}\\
0&0&\eta_{\rm R}&(1+\delta)-\eta_{\rm L}
\end{array}
\right]_{jj+1}\nonumber\\
&=&I+\Delta H_{\rm ASEP},
\nonumber\\
\Delta H_{\rm ASEP}
&=&\frac{2}{N}\sum_{j=1}^N[
 \frac{1}{2}(\eta_{\rm R}s_j^-s_{j+1}^++\eta_{\rm L}s_j^+s_{j+1}^-)
+\frac{1}{2}(\eta_{\rm R}+\eta_{\rm L})s_j^zs_{j+1}^z
\nonumber\\
&&\hspace{0.8cm}-\frac{1}{8}(\eta_{\rm R}+\eta_{\rm L})I
+\frac{1}{4}(\eta_{\rm L}-\eta_{\rm R}-2\delta)(s_j^z-s_{j+1}^z)].
\label{ASEP2}
\end{eqnarray}
This is the ASEP with discretized time step. 
Because of the periodic boundary condition, 
the last term vanishes and the operator is independent of $\delta$. 
The rates $\delta$ and $-\delta$ always appear pairwise 
in the summation $\sum_{j=1}^N$, 
and finally give no contribution to the probability. 
When one set $\eta_{\rm R}=1-\delta$ and $\eta_{\rm L}=1+\delta$, 
the system is reduced to (\ref{RW1}) and (\ref{RW2}). 
In this case, 
the ASEP is a simple random walk process with hard-core interactions. 

As a model of the movement of ribosomes, 
the rates are asymmetric and the open boundary condition should be introduced.   
With these restrictions, the expression of $H_{\rm ASEP}$ by spin operators 
have already been written by Sandow (1994, see also Alcaraz 1994). 
The condition (\ref{condstoch3}) is satisfied provided $\delta=0$. 
The attachment at the site $j=1$ and the displacement at the site $j=N+1$ 
are introduced by the terms 
$\alpha_1[s_1^+-(1-n_1)]$ and $\beta_{N+1}[s_{N+1}^--n_{N+1}]$. 
Then with the use of the transformation 
$V=\exp[(\log q)\sum_{j=1}^{N+1} jn_j]$, 
which is slightly modefied from (\ref{transV}), one obtains 
\begin{eqnarray}
V\Delta H_{\rm ASEP}V^{-1}
&=&\frac{2}{N}\sqrt{\eta_{\rm R}\eta_{\rm L}}
\sum_{j=1}^N[
s_j^xs_{j+1}^x+s_j^ys_{j+1}^y+\frac{1}{2}(q+q^{-1})s_j^zs_{j+1}^z
-\frac{1}{8}(q+q^{-1})I]
\nonumber\\
&&+\frac{1}{N}\sqrt{\eta_{\rm R}\eta_{\rm L}}\frac{1}{2}(q-q^{-1})(s_1^z-s_{N+1}^z)
\nonumber\\
&&+\alpha_1[qs_1^+-(1-n_1)]+\beta_{N+1}[q^{-(N+1)}s_{N+1}^--n_{N+1}].
\nonumber
\end{eqnarray}
Therefore the ASEP with asymmetric rates and with the open boundary condition 
is nothing but the XXZ spin chain with $\Delta=(q+q^{-1})/2$ 
with boundary magnetic fields. 
Here we find that $q$ is the parameter 
in the quantum group symmetry of the XXZ spin chain 
(Pasquier and Saleur 1990; Jimbo and Miwa 1993).

Other generalizations of the ASEP reviewed above can be handled 
as generalizations of spin Hamiltonians. 
The biochemical states can be introduced using spin operators with the spin magnitude $S$, 
in which the number of the eigenstates is $n=2S+1$, 
and the bottleneck or the non-uniform hopping rates 
correspond to non-uniform coupling constants of spin chains. 
The ASEP with continuous time variation 
is also governed by the same operator (\ref{ASEP2}): 
its time dependence is 
\begin{eqnarray}
\frac{d}{dt}{\bf P}(t)=(\Delta H_{\rm ASEP}){\bf P}(t), 
\nonumber
\end{eqnarray}
where each element of ${\bf P}(t)$ is equal to 
$P(x_1, x_2, \ldots, x_{m_{\rm p}}; t)$, 
the probability that the system is in the configuration 
$\{x_1, x_2, \ldots, x_{m_{\rm p}}\}$ at time $t$. 
\newline

\noindent
{\bf Kinesins moving on a microtube:}
Kinesin is a single molecular motor observed in vitro 
to move along a linear microtubule template. 
It moves stochastically and stepwise along a one-dimensional track  
(for example, a review by Yildiz and Selvin 2005). 

Experimental methods to measure 
the biochemical and biomechanical properties of a single-molecule 
enable us to observe the movement of a single kinesin. 
It was shown that kinesin moves stepwise along microtubules 
occasionally moving both forwards and backwards 
(Kojima et al. 1997). 
It was observed that 
kinesin is released spontaneously from the microtubule 
(Block et al. 1990), and 
that the increase of the load results in an increasing rate of dissociation 
(Coppin et al. 1997).
It was suggested that two or more sequential processes dominate the biochemical cycle 
(Svoboda et al. 1994), and 
the time to force generation after release of ATP  was measured 
(Higuchi et al. 1997). 
Force-velocity curves was obtained for single kinesin molecules 
(Svoboda and Block 1994).

Pioneering theoretical models related to the dynamics of  kinesin are already known. 
An elementary "barometric" relation was introduced for the driving force 
(Fisher and Kolomeisky 1999a; b). 
Nearest-neighbor kinetic hopping models 
with arbitrary forward and backward periodic rate constants 
with three generalizations was introduced 
(Kolomeisky and Fisher 2000a). 
The one-dimensional random walk process with general waiting-time distributions, 
finite side branches, and annihilation was also considered 
(Kolomeisky and Fisher 2000b).
The observed movement of kinesin can be described adequately 
by simple discrete-state stochastic models 
(Fisher and Kolomeisky 2001). 
The model has also been applied to analyze the dynamics of myosin-V 
(Kolomeisky and Fisher 2003).   
Brownian particles moving in two or more periodic but spatially asymmetric and stochastically switched potentials have also been considered 
(J\"{u}licher et al. 1997) 
The models have been summarized in review articles 
(Reimann 2002; Kolomeisky and Fisher 2007).

Let us consider a model for kinesin on a microtubel 
(Kolomeisky and Fisher 2000a). 
A kinesin is assumed to locate on a site $j$ of a one-dimensional lattice, 
where $j=1,2, \ldots, N+1$. 
Let us introduce an index $k\;\;(k =0, 1, 2, \ldots, K-1)$ 
which distinguishes the internal states of kinesin.  
A kinesin changes its internal state from $k$ to $k+1$ with the rate $u_k$, 
and from $k$ to $k-1$ with the rate $w_k$. 
A kinesin at site $j$ in the maximum internal state $k=K-1$ 
can move to the next site $j+1$ and initialize its state as $k=0$
with the rate $u_{K-1}$.  
A kinesin at site $j+1$ in the minimum state $k=0$ 
can move to the previous site $j$ and have the maximum state $k=K-1$ 
with the rate $w_0$. 

Let us consider the case with two internal states: the case $K=2$. 
Each site takes one of the three possible states: $k=0$, $1$, and  "empty". 
Thus let us introduce the spin operator with the spin magnitude $S=1$,  
in which there are $2S+1=3$ possible eigenstates with the eigenvalues 
$S^z=-1$, $+1$, and $0$. 
Let us assume that the states with $S^z=-1$, $+1$, and $0$ 
correspond to the states $k=0$, $1$, and  "empty", respectively. 
Let the generating matrix be $H_{\rm KF}=I+\Delta H_{\rm KF}$, 
and consider $\Delta H_{\rm KF}$. 
The hopping from the state $k=0$ to $k=1$ at site $j$ is expressed as 
\begin{eqnarray}
H_j^+&=&u_0
\left(
\left[
\begin{array}{ccc}
0&0&1\\
0&0&0\\
0&0&0
\end{array}
\right]_{j}
-
\left[
\begin{array}{ccc}
0&0&0\\
0&0&0\\
0&0&1
\end{array}
\right]_{j}
\right)
\nonumber\\
&=&
u_0[(\frac{1}{\sqrt{2}}s_j^+)^2-(I_j-\frac{1}{\sqrt{2}}s_j^+\frac{1}{\sqrt{2}}s_j^-)]
=u_0(s_j^+s_j^x-I_j).
\nonumber
\end{eqnarray}
The backward hopping from the state $k=1$ to $0$ is, in the same way, expressed as 
$H_j^-=w_1(s_j^-s_j^x-I_j)$. 
The hopping from the state $k=1$ at site $j$ 
to the state $k=0$ at site $j+1$ is expressed as 
\begin{eqnarray}
H_{jj+1}^+&=&u_1
\left(
\left[
\begin{array}{ccc}
0&0&0\\
1&0&0\\
0&0&0
\end{array}
\right]_{j}
\otimes
\left[
\begin{array}{ccc}
0&0&0\\
0&0&0\\
0&1&0
\end{array}
\right]_{j+1}
-
\left[
\begin{array}{ccc}
1&0&0\\
0&0&0\\
0&0&0
\end{array}
\right]_{j}
\otimes
\left[
\begin{array}{ccc}
0&0&0\\
0&1&0\\
0&0&0
\end{array}
\right]_{j+1}
\right)
\nonumber\\
&=&
u_1\left[
(\frac{1}{\sqrt{2}}s_j^-s_j^z)(s_{j+1}^z\frac{-1}{\sqrt{2}}s_{j+1}^-)
-(I_j-\frac{1}{\sqrt{2}}s_j^-\frac{1}{\sqrt{2}}s_j^+)(I_{j+1}-(s_{j+1}^z)^2)
\right]
\nonumber\\
&=&
-\frac{1}{2}u_1
\left[
(s_j^-s_j^z)(s_{j+1}^zs_{j+1}^-)
+((s_j^z)^2+s_j^z)(I_{j+1}-(s_{j+1}^z)^2)
\right],
\nonumber
\end{eqnarray}
where we made use of the relations 
$(s_j^x)^2+(s_j^y)^2+(s_j^z)^2=S(S+1)=2$ and 
\begin{eqnarray}
I_j-\frac{1}{\sqrt{2}}s_j^\pm\frac{1}{\sqrt{2}}s_j^\mp
=\frac{1}{2}[(s_j^z)^2\mp s_j^z].
\nonumber
\end{eqnarray}
The backward hopping from the state $k=0$ at site $j+1$ 
to the state $k=1$ at site $j$ is, in the same way, expressed as 
\begin{eqnarray}
H_{jj+1}^-=
-\frac{1}{2}w_0
\left[
(s_j^zs_j^+)(s_{j+1}^+s_{j+1}^z)
+(I_j-(s_j^z)^2)((s_{j+1}^z)^2-s_{j+1}^z)
\right].
\nonumber
\end{eqnarray}
These operators constitute $\Delta H_{\rm KF}$. 
The operator $H_{\rm KF}$ is then obtained as 
\begin{eqnarray}
H_{\rm KF}=I+\sum_{j=1}^{N+1}( H_j^++H_j^-)+\sum_{j=1}^N(H_{jj+1}^++H_{jj+1}^-).   
\nonumber
\end{eqnarray}
The operator should be normalized by the maximum eigenvalue $\lambda_0$. 
\newline

\noindent
{\bf General correspondence:}
In these three examples 
the generation of the biological stochastic systems 
are written in terms of the spin Hamiltonians. 
The correspondence between stochastic systems and spin models are more general. 
Let us consider a system with $n$ discrete internal states. 
The law governing changes from one state to another 
is written as a matrix of order $n$. 
Let us derive the fact that 
this matrix can always be expressed by the spin operators with spin $S$, where $n=2S+1$. 

Let us consider the unit matrix $I$ and the matrices $(s^z)^k\:\:(k\in{\bf N})$. 
They are diagonal and the diagonal elements of $(s^z)^k$ are 
$S^k$, $(S-1)^k$, \ldots, $(-(S-1))^k$ and $(-S)^k$.  
The set of the matrices $I, s^z, (s^z)^2, \ldots, (s^z)^{2S}, (s^z)^{2S+1}$  
are not independent because they satisfy the eigenequation of $s^z$, 
which is a polynomial of order $2S+1$.  
For the purpose to show the independence of 
$I, s^z, (s^z)^2, \ldots, (s^z)^{2S}$, 
let us consider the following determinant
\begin{eqnarray}
&\det&
\left(
\begin{array}{cccccc}
1&S & S^2&\cdots &S^{2S} \\
1&S-1 &(S-1)^2 & & (S-1)^{2S}\\
1&\vdots & & & \\
1&-(S-1) &(-(S-1))^2 && (-(S-1))^{2S}  \\
1&-S &(-S)^2 &\cdots & (-S)^{2S} \\
\end{array}
\right)
\nonumber\\ 
\nonumber\\
&=&(-1)^{\frac{1}{2}n(n-1)}\prod_{1\leq i< j\leq n}(z_i-z_j), 
\nonumber
\end{eqnarray}
where $n=2S+1$ and 
\begin{eqnarray*}
z_1=S, \:\: z_2=S-1, \:\ldots\: , \:\: z_{2S+1}=-S. 
\nonumber
\end{eqnarray*}
This is the Vandermonde's determinant and, in our case, clearly non-zero. 
Hence all the diagonal matrices can be expressed 
as a linear combination of $I$ and $(s^z)^k\:\:(k=1, 2, \ldots, 2S+1)$.

In the case of spin operators with the spin magnitude $S$, 
it is derived from the commutation relations 
that the eigenstate of $s^z$ corresponding to the eigenvalue $M$ satisfies 
$s^\pm|M\rangle=[S(S+1)-M(M\pm 1)]^{1/2}|M\pm 1\rangle$.   
Thus the matrix elements of $s^\pm$ satisfy 
$(s^\pm)_{ij}\neq 0$ if and only if $j=i\pm 1$, 
and $((s^\pm)^l)_{ij}\neq 0$ if and only if $j=i\pm l$.  
 Let us introduce a diagonal matrix $P_k$ by $(P_k)_{ij}=\delta_{ki}\delta_{kj}$, 
i.e. $(P_k)_{kk}=1$ and $(P_k)_{ij}=0$ when $i\neq k$ or $j\neq k$. 
It is easy to convince that 
\begin{eqnarray}
(P_k(s^+)^l)_{ij}&\neq& 0\hspace{0.6cm}i=k, \:\:j=k+l, \nonumber\\
(P_k(s^-)^l)_{ij}&\neq& 0\hspace{0.6cm}i=k, \:\:j=k-l, 
\nonumber
\end{eqnarray}
and all the other matrix elements are equal to $0$. 
Therefore arbitrary $n\times n$ matrices can be expressed 
in terms of the spin operators with spin $S$, where $n=2S+1$. 

The Hamiltonian generally has the form 
\begin{eqnarray}
H
=\sum_{\langle i,j\rangle}\sum_{\alpha\alpha'}
J_{mn}^{(\alpha\alpha')}(s_i^\alpha)^m(s_j^{\alpha'})^n,
\label{spinHgen}
\end{eqnarray}
where $\alpha$ and $\alpha'$ denotes $z$, $+$, or $-$. 
An external field can be introduced 
through the terms with the powers $(m, n)=(0, 1)$ or $(1, 0)$. 
Therefore 
all the biological systems with finite number of states and discrete law of change 
can find its equivalent spin model.

\section{Summary and discussions}
In this study, it was derived that 
the adhesion probabilities in a two-dimensional cell-sorting model 
are analytically expressed by the expectation values 
in the one-dimensional random walk model with pair creations and annihilations. 
I believe that this is the first example 
in which completely different biological systems 
show mathematical equivalence to each other. 
It was also derived that 
the equivalencies with spin systems are general, 
i.e. the generation rules of stochastic movements in biological systems 
can always be written in terms of the lattice spin Hamiltonians. 
This provides a path to study biological systems 
using the techniques and results 
already obtained in the area of spin systems. 

It should be noted that 
general correspondences between $d$-dimensional quantum spin systems 
and $(d+1)$-dimensional Ising-type spin systems are known to exist 
(Suzuki 1976). 
This implies that 
for each stochastic system in $d$-dimension 
with hopping, creation and annihilation of particles, 
there exist equivalent $(d+1)$-dimensional model 
with an adhesion-type structure. 

Although the stochastic models can be expressed in terms of spin operators, 
the corresponding Hamiltonians are usually not simple. 
In such situations, numerical calculation techniques might be powerful, 
and many sophisticated techniques 
have been developed for the lattice spin models 
(see for example the series by Domb and Green 1972) .  
Exact results also exist, for example, in the case of $\alpha=\alpha'=z$ 
in the one-dimensional case of (\ref{spinHgen}), 
that are $n$-state Ising-type systems with independent one-particle creations and annihilations 
(Minami 1998).  

The dynamics of the Ising model have been studied through the system 
proposed by Glauber (Glauber 1963) or Kawasaki (Kawasaki 1966). 
The cell-sorting model considered in this study with pair exchange probability 
corresponds to the Kawasaki dynamics. 
The dynamics of $n$-spin flip models were considered, 
in which the two-spin flip-model is equivalent to the XY chain 
apart from the boundary terms 
(Felderhof and Suzuki 1971). 
It is interesting to consider the equivalencies of dynamical properties. 

The XXZ model is related to the random walk model, 
and therefore related to the Brownian motion. 
The one-dimensional XXZ model is equivalent to the six-vertex model, 
and it is known to demonstrate fractal structure in its configuration space 
(Minami 2010). 
The XY model with periodically varying interactions also becomes interesting 
from biological viewpoints. \\

This work is partially supported by a Grant-in-Aid for Scientific Research  
by the Ministry of Education, Culture, Sports, Science and Technology, Japan. 

\newpage

\noindent
{\large\bf References}
\vspace{0.6cm}

\noindent
Antonelli  P,  McLaren  DI,  Rogers  TD,  Lathrop  M, Willard  MA  (1975)
Transitivity  pattern-reversal, engulfment and duality in exchange-type cell aggregation kinetics. 
J Theor Biol 49:385-400 
\\

\noindent
Antonelli  PL,  Rogers  TD,  Willard  MA (1973)
Geometry and exchange principle in cell aggregation kinetics. 
J Theor Biol 41:1-21 
\\

\noindent
Alcaraz  FC, Droz M, Henkel M (1994)
Reaction-diffusion processes, critical dynamics, and quantum chains. 
Ann Phys 230:250-302 
\\

\noindent
Basu A, Chowdhury D (2007) 
Traffic of interacting ribosomes: Effects of single-machine mechanochemistry on protein synthesis. 
Phys Rev E 75:021902 
\\

\noindent
Block SM, Goldstein LSB, Schnapp BJ (1990)
Bead movement by single kinesin molecules studied with optical tweezers. 
Nature 348:348-352 
\\

\noindent
Chou T (2003)
Ribosome recycling, diffusion, and mRNA loop formation in translational regulation. 
Biophys J 85:755-773 
\\

\noindent
Chou T, Lakatos G (2004)
Clustered bottlenecks in mRNA translation and protein synthesis. 
Phys Rev Lett 93:198101 
\\

\noindent
Ciandrini L, Stanseld I, Romano MC (2010)
Role of the particle's stepping cycle in an asymmetric exclusion process: A model of mRNA translation. 
arXiv:0912.3482v2 
\\

\noindent
Coppin CM, Pierce DW, Hsu L, Vale RD (1997)
The load dependence of kinesin's mechanical cycle. 
Proc Natl Acad Sci USA 94:8539-8544 
\\

\noindent
Derrida B (1998) 
An exactly soluble non-equilibrium system: The asymmetric simple exclusion process. 
Phys Rep 301:65-83 
\\

\noindent
Domb C, Green MS (1972) 
Phase transitions and critical phenomena. 
Academic Press
\\

\noindent
Dong JJ, Schmittmann B, Zia RKP  (2007)
Inhomogeneous exclusion processes with extended objects: The effect of defect locations. 
Phys Rev E 76:051113 
\\

\noindent
Dong JJ, Schmittmann B, Zia RKP (2007)
Towards a model for protein production rates. 
J Stat Phys 128:21-34 
\\

\noindent
Felderhof BU, Suzuki M (1971)
Time-correlation functions and critical relaxation in a class of one-dimensional stochastic spin systems. 
Physica 56:43-61 
\\

\noindent
Fisher ME (1960) 
The perpendicular susceptibility of an anisotropic antiferromagnet. 
Physica 26:618-622 
\\

\noindent
Fisher ME (1963)
Perpendicular susceptibility of the Ising model. 
J Math Phys 4:124-135
\\

\noindent
Fisher ME, Kolomeisky AB  (1999)
Molecular motors and the forces they exert. 
Physica A 274:241-266
\\

\noindent
Fisher ME, Kolomeisky AB (1999)
The force exerted by a molecular motor. 
Proc Natl Acad Sci USA 96:6597-6602
\\

\noindent
Fisher ME, Kolomeisky AB (2001)
Simple mechanochemistry describes the dynamics of kinesin molecules. 
Proc Natl Acad Sci USA 98:7748-7753
\\

\noindent
Garai A, Chowdhury D, Chowdhury D, Ramakrishnan TV (2009) 
Stochastic kinetics of ribosomes: Single motor properties and collective behavior. 
Phys Rev E 80:011908
\\

\noindent
Glauber RJ (1963)
Time]dependent statistics of the Ising model. 
J Math Phys 4:294-307
\\

\noindent
Glazier  JA, Graner  F (1993)
Simulation of the differential adhesion driven rearrangement of biological cells. 
Phys Rev E47:2128-2154 
\\

\noindent
Goel  NS,  Campbell  RD,  Gordon  R,  Rosen  R, Martinez  H,  Y\v{c}as  M (1970)
Self-sorting of isotropic cells. 
J Theor Biol 28:423-468
\\

\noindent
Goel  NS,  Leith AG (1970)
Self-sorting of anisotropic cells. 
J Theor Biol 28:469-482
\\

\noindent
Goel  NS,  Rogers  G  (1978)
Computer simulation of engulfment and other movements of embryonic tissues. 
J Theor Biol 71:103-140
\\

\noindent
Gordon  R,  Goel  NS,  Steinberg  MS,  Wiseman  LL (1972)
A rheological mechanism sufficient to explain the kinetics of cell sorting. 
J Theor Biol 37:43-73 
\\

\noindent
Graner  F, Glazier  JA (1992) 
Simulation of biological cell sorting using a two-dimensional extended Potts model. 
Phys Rev Lett 69:2013-2016
\\

\noindent
Graner  F, Sawada  Y  (1993)
Can surface adhesion drive cell-rearrangement?  Part II: A geometrical model. 
J Theor Biol 164:477-506
\\

\noindent
Greenspan  D (1981) 
A classical molecular approach to computer simulation of biological sorting. 
J Math Biol 12:227-235 
\\

\noindent
Henkel H, Sch\"{u}tz G (1994)
Boubdary-induced phase transitions in equilibrium and non-equilibrium systems. 
Physica A 206:187-195
\\

\noindent
Higuchi H, Muto E, Inoue Y, Yanagida T (1997)
Kinetics of force generation by single kinesin molecules activated by laser photolysis of caged ATP. 
Proc Natl Acad Sci USA 94:4395-4400
\\

\noindent
Ising E (1925)
Report on the theory of ferromagnetism. 
Zeits F Phys 31:253-258
\\

\noindent
Jimbo M, Miwa T (1993)
Algebraic analysis of solvable lattice models. 
CBMS Regional Conference Series in Mathematics 85, AMS
\\

\noindent
J\''{u} licher F, Ajdari A, Prost J (1997)
Modeling molecular motors. 
Rev Mod Phys 69:1269-1281
\\

\noindent
Katsura S (1962) 
Statistical mechanics of the anisotropic linear Heisenberg model. 
Phys Rev 127:1508-1518
\\

\noindent
Kawasaki K (1966)
Diffusion constants near the critical point for time-dependent Ising models. I. 
Phys Rev 145:224-230
\\

\noindent
Kojima H, Muto E, Higuchi H, Yanagida T (1997)
Mechanics of single kinesin molecules measured by optical trapping nanometry. 
Biophys J 73:2012-2022
\\

\noindent
Kolomeisky AB (1998)
Asymmetric simple exclusion model with local inhomogeneity. 
J Phys A Math Gen 31:1153-1164
\\

\noindent
Kolomeisky AB, Fisher ME (2000)
Periodic sequential kinetic models with jumping, branching and deaths. 
Physica A 279:1-20
\\

\noindent
Kolomeisky AB, Fisher ME (2000)
Extended kinetic models with waiting-time distributions: Exact results. 
J Chem Phys 113:10867
\\

\noindent
Kolomeisky AB, Fisher ME (2003)
A simple kinetic model describes the processivity of myosin-V. 
Biophys J 84:1642-1650
\\

\noindent
Kolomeisky AB, Fisher ME (2007)
Molecular motors: A theoristfs perspective. 
Ann Rev Phys Chem 58:675-695
\\

\noindent
Kramers HA, Wannier GH  (1941) 
Statistics of the two-dimensional ferromagnet. Part I. 
Phys Rev 60:252-262 
\\

\noindent
Kubo R  (1943)
An analytic method in statistical mechanics.  
Busseiron Kenkyu 1:1-13
\\

\noindent
Lieb EH, Schults TD, Mattis DC (1961) 
Two soluble models of an antiferromagnetic chain. 
Ann Phys 16:407-466
\\

\noindent
Leith  AG,  Goel  NS (1971)
Simulation of movement of cells during self-sorting. 
J Theor Biol 33:171-188
\\

\noindent
MacDonald CT, Gibbs JH (1969) 
Concerning the kinetics of polypeptide synthesis on polyribosomes. 
Biopolymers 7:707-725 
\\

\noindent
MacDonald CT, Gibbs JH, Pipkin AC (1968) 
Kinetics of biopolymerization on nucleoc acid templates. 
Biopolymers 6:1-25
\\

\noindent
Matela  RJ,  Fletterick  RJ (1979)
A topological exchange model for cell self-sorting. 
J Theor Biol 76:403-414
\\

\noindent
Matela  RJ, Fletterick  RJ (1980)
Computer simulation of cellular self-sorting: A topological exchange model. 
J Theor Biol 84:673-690
\\

\noindent
Matsubara T, Matsuda H (1956)
A lattice model of liquid helium, I. 
Prog Theor Phys 16:569-582
\\

\noindent
McCoy BM, Barouch E, Abraham DB (1971)
Statistical mechanics of the XY model. IV.  Time-dependent spin-correlation functions. 
Phys Rev A 4:2331-2341
\\

\noindent
McCoy BM, Tracy CA, Wu TT (1977)
Two-dimensional Ising model as an exactly solvable relativistic quantum field theory: Explicit formulas for n-point functions. 
Phys Rev Lett 38:793-796
\\

\noindent
Minami K (1996)
The zero-field susceptibility of the transverse Ising chain with arbitrary spin. 
J Phys A Math Gen 29:6395-6405
\\

\noindent
Minam K (1998)
The susceptibility in arbitrary directions and the specific heat in general Ising-type chains of uniform, periodic and random structures. 
J Phys Soc Jpn 67:2255-2269
\\

\noindent
Minami K (2010)
Fractal structure of a solvable lattice model. 
Int J Pure App Math 59:243-255
\\

\noindent
Mochizuki A (2002)
Pattern formation of the cone mosaic in the zebrafish retina: A cell rearrangement model. 
J Theor Biol 215:345-361
\\

\noindent
Mochizuki A, Iwasa Y,  Takeda Y (1996)
A stochastic model for cell sorting and measureing cell-cell adhesion. 
J Theor Biol 179:129-146
\\

\noindent
Mochizuki A, Wada N, Ide H, Iwasa Y (1998)
Cell-cell adhesion in limb-formation, estimated from photographs of cell sorting experiments based on a spatial stochastic model. 
Dev  Dynamics 211:204-214
\\

\noindent
Nakajima A, Ishihara S (2011)
Kinetics of the cellular Potts model revisited. 
New J Phys 13:033035 
\\

\noindent
Niemaijer T (1967)
Some exact calculations on a chain of spin 1/2. 
Physica 36:377-419 
\\

\noindent
Onsager L (1944) 
Crystal statistics. I. A two-dimensional model with an order-disorder transition. 
Phys Rev 65:117-149 
\\

\noindent
Parmeggiani A, Franosch T, Frey E (2003) 
Phase coexistence in driven one-dimensional transport. 
Phys Rev Lett 90:086601
\\

\noindent
Pasquier V, Saleur H (1990) 
Common structures between finite systems and conformal field theories through quantum groups. 
Nuc Phys B330:523-556
\\

\noindent
Pfeuty P (1970) 
The one-dimensional Ising model with a transverse field. 
Ann Phys 57:79-90 
\\

\noindent
Pierobon P, Mobilia M, Kouyos R, Frey E (2006) 
Bottleneck-induced transitions in a minimal model for intracellular transport. 
Phys Rev E 74:031906 
\\

\noindent
Reimann P (2002) 
Brownian motors: noisy transport far from equilibrium. 
Phys Rep 361:57-265
\\

\noindent
Rogers  G,  Goel  NS  (1978) 
Computer simulation of cellular movements: Cell-sorting, cellular migration through a mass of cells and contact inhibition. 
J Theor Biol 71:141-166
\\

\noindent
Rogers TD, Sampson JR (1977) 
A random walk model of cellular kinetics. 
Int J Bio-Med Comp 8:45-60
\\

\noindent
Romano MC, Thiel M, Stansfield I, Grebogi C (2009) 
Queueing phase transition: Theory of translation. 
Phys Rev Lett 102:198104 
\\

\noindent
Sandow S (1994) 
Partially asymmetric exclusion process with open boundary. 
Phys Rev E 50:2660-2667
\\

\noindent
Shaw LB, Sethna JP, Lee KH (2004) 
Mean-field approaches to the totally asymmetric exclusion process with quenched disorder and large particles. 
Phys Rev E 70:021901
\\

\noindent
Shaw LB, Zia RKP, Lee KH (2003) 
Totally asymmetric exclusion process with extended objects: A model for protein synthesis. 
Phys Rev E 68:021910 
\\

\noindent
Steinberg  MS  (1962) 
On the mechanism of tissue reconstruction by dissociated cells, I. Population kinetics, differential adhesiveness, and the absence of directed migration.  
Proc Natn Acad Sci USA 48:1577-1582
\\

\noindent
Steinberg  MS  (1962) 
Mechanism of tissue reconstruction by dissociated cells II: Time-course of events.  
Science 137:762-763
\\

\noindent
Steinberg  MS  (1962) 
On the mechanism of tissue reconstruction by dissociated cells, III. Free energy relations and the reorganization of fused, heteronomic tissue fragments. 
Proc Natn Acad Sci USA 48:1769-1776
\\

\noindent
Steinberg  MS  (1963) 
Reconstruction of tissues by dissociated cells. 
Science 141: 401-408
\\

\noindent
Steinberg  MS  (1970) 
Does differential adhesion govern self-assembly processes in histogenesis? Equilibrium configurations and the emergence of a hierarchy among populations of embryonic cells.  
J Exp Zool 173:395-434
\\

\noindent
Sulsky  D,  Childress  S,  Percus  JK  (1984) 
A model of cell sorting. 
J Theor Biol 106:275-301
\\

\noindent
Suzuki M (1971) 
Relationship among exactly soluble models of critical phenomena I. 
Prog Toer Phys 46:1337-1359
\\

\noindent
Suzuki M (1976) 
Relationship between d-dimensional quantal spin systems and (d+1)-dimensional Ising system : Equivalence, critical exponents and systematic approximants of the partition function and spin correlations. 
Prog Theor Phys 56:1454-1469
\\

\noindent
Svoboda K, Block SM (1994)
Force and velocity measured for single kinesin molecules. 
Cell 77:773-794
\\

\noindent
Svoboda K, Mitra PP, Block SM (1994) 
Fluctuation analysis of motor protein movement and single enzyme kinetics. 
Proc Natl Acad Sci USA 91:11782-11786
\\

\noindent
Tonegawa T (1981)
Transverse spin correlation function of the one-dimensional spin-1/2 XY model. 
Solid State Comm 40:983-986
\\

\noindent
Umeda T  (1989)
A mathematical model for cell-sorting, migration and shape in the slug state of dictyostelium discoideum. 
Bull Math Biol 51:485-500
\\

\noindent
Umeda T,  Inouye K (1999)
Theoretical model for morphogenesis and cell sorting in dictyostelium discoideum. 
Physica D 126:189-200
\\

\noindent
Umeda T, Inouye K (2004)
Cell sorting by differential cell mobility: A model for pattern formation in dictyostelium. 
J Theor Biol 226:215-224 
\\

\noindent
von Heijine G, Nilsson L, Blomberg C (1977) 
Translation and messenger RNA secondary structure. 
J Theor Biol 68:321-329
\\

\noindent
Wu TT, McCoy BM, Tracy CA, Barouch E  (1976)
Spin-spin correlation functions for the two-dimensional Ising model: Exact theory in the scaling region. 
Phys Rev B13:316-374
\\

\noindent
Yang CN (1952) 
Spontaneous magnetization of a two-dimensional Ising model. 
Phys Rev 85:808-816 
\\

\noindent
Yildiz A, Selvin PR (2005)
Kinesin: Walking, crawling or sliding along?
Trends Cell Biol 15:112-120
\\

\end{document}